\documentclass[aps,prb,twocolumn,groupedaddress]{revtex4}
\usepackage{graphicx}

\begin{document}

\title{Numerical estimation of the $\beta$-function
in 2D systems with spin-orbit coupling}

\author{Yoichi Asada and Keith Slevin}
\affiliation{Department of Physics, Graduate School of Science, 
Osaka University, 1-1 Machikaneyama, Toyonaka, Osaka 560-0043, Japan}

\author{Tomi Ohtsuki}
\affiliation{Department of Physics, 
Sophia University, Kioi-cho 7-1, Chiyoda-ku, Tokyo 102-8554, Japan}

\date{\today}

\begin{abstract}
We report a numerical study of Anderson localization in a 2D system of non-interacting
electrons with spin-orbit coupling.
We analyze the scaling of the renormalized localization length
for the 2D SU(2) model and estimate its 
$\beta$-function over the
full range from the localized to the metallic limits.
\end{abstract}

\pacs{}

\maketitle

\section{Introduction}

As a general rule, all states in a disordered two dimensional (2D) system of
non-interacting electrons are localized.\cite{abrahams:79}
There are two exceptions.
One is the quantum Hall effect which occurs in
2D systems subject to strong perpendicular magnetic fields,
where delocalized states exist at the center of a Landau level.
\cite{huckestein:95}
Another is the metallic phase that
occurs in 2D systems with symplectic symmetry, i.e.
in systems with time reversal symmetry but in which spin
rotation symmetry is broken.\cite{hikami:80,ando:89}
The latter of these, which is the subject of this paper,
is realized when the spin-orbit interaction is significant:
this interaction breaks spin rotation
symmetry but not time reversal symmetry.

While progress has been made,\cite{kawabata:88,grempel:87}
a satisfactory analytic theory of the 
metal-insulator transition
in 2D systems with symplectic symmetry
has yet to be developed.
In the absence of such a theory, 
numerical simulation remains the most useful tool with which to investigate
the transition.\cite{evangelou:87,ando:89,fastenrath:91}

In a recent Physical Review Letter\cite{asada:02-1} we estimated the
critical exponent for this transition.
Prior to our work varying estimates of the critical exponent
had been reported.
The main obstacle to a higher precision estimate of the exponent
had been corrections to scaling due to irrelevant scaling
variables in many of the models analyzed numerically.
We overcame this difficulty by proposing an SU(2) model
for which such corrections are much less significant.

In this paper we present a detailed analysis of the scaling of the
quasi-1D localization length in the SU(2) model in the metallic,
critical and localized regimes.
Scaling in the critical and localized regimes was not dealt with in our
Letter.
From this we have estimated the renormalization
group $\beta$-function for the quasi-1D localization length.
We also present supplementary results for the phase diagram and rule
out the occurrence of re-entrant behavior.

\section{Model and method}

\subsection{SU(2) model}

We simulated the SU(2) model \cite{asada:02-1}
\begin{eqnarray}
H=\sum_{i,\sigma}\epsilon_i c_{i\sigma}^{\dagger}c_{i\sigma}
-\sum_{\langle i,j \rangle,\sigma,\sigma'}
R(i;j)_{ \sigma \sigma'}
c_{i\sigma}^{\dagger}c_{j\sigma'}
\label{eq:hamiltonian}
\end{eqnarray}
where $c_{i\sigma}^{\dagger}(c_{i\sigma})$ denotes
the creation (annihilation) operator
of an electron at the site $i=(x,y)$ with spin $\sigma$.
The random potential $\epsilon_i$
is identically and independently distributed with uniform probability
on $[-W/2,W/2]$.
Hopping is restricted to nearest neighbors
and the hopping matrix $R(i;j)$ is distributed
randomly and independently with uniform probability
on the group SU(2) according to the group
invariant measure.
More explicitly
\begin{equation}
	R(i;j)= \left(
\! \!
\begin{array}{cc}
e^{i\alpha_{ij}} \cos \beta_{ij}
& e^{i\gamma_{ij}} \sin \beta_{ij} \\
-e^{-i\gamma_{ij}} \sin \beta_{ij}
& e^{-i\alpha_{ij} }\cos \beta_{ij}
\end{array}
\! \!
\right)
\end{equation}
with $\alpha_{ij}$ and $\gamma_{ij}$ uniform on $\left[0,2\pi\right)$, and
$\beta_{ij}$ distributed according to
\begin{equation}
P(\beta) \rm d \beta = 
\left\{
\begin{array}{ll}
\sin (2\beta) \rm d \beta & 0 \le \beta \le \frac{\pi}{2} \\ 
 0                                  & \rm otherwise.
\end{array}
\right.
\end{equation}
We perform an SU(2) gauge transformation on the model
for reasons of numerical efficiency (see Appendix \ref{app:gaugetrans}).

\subsection{Transfer matrix method}

We have used two different variants of
the transfer matrix method to estimate the localization length
$\lambda$ of an electron on quasi-1D strips with transverse
dimension $L_t$ and length $L_z$. 
We impose periodic boundary conditions in the transverse
direction.
We use quaternion arithmetic to perform the transfer matrix calculations.

The first traditional transfer matrix method\cite{mackinnon:83}
estimates the localization length by simulating a single very long
sample $L_z\gg L_t$.
The length of the sample is increased until a desired precision for 
$\lambda$ is obtained.

The second method\cite{slevin:04} called here the ensemble transfer matrix method
simulates an ensemble of samples with a fixed length $L_z$.
Here the number of samples is increased until a desired precision for
the localization length is achieved.
To ensure that the estimate of the localization length is 
independent of $L_z$, a special choice of the distribution of
the starting vectors in the transfer matrix iteration is required.
We take a set of ortho-normal vectors and perform $N_r$ transfer matrix
iterations on these vectors with Gram-Schmidt orthogonalizations.
If $N_r$ is sufficiently large, a stationary distribution of ortho-normal vectors is reached.
When vectors sampled from this stationary distribution are used as starting vectors, the estimate of
the localization length becomes independent of $L_z$.

\subsection{Finite size scaling method}
\label{sec:fssmethod}

Our analysis is based on the assumption that the 
renormalized localization length $\Lambda$, defined by
\begin{equation}
	\Lambda=\frac{\lambda}{L_t},
	\label{eq:defLambda}
\end{equation}
obeys single parameter scaling (SPS) law that is described by the $\beta$-function\cite{mackinnon:83}
\begin{equation}
\beta\left(\ln\Lambda\right)=
\frac{{\mathrm d}\ln\Lambda}{{\mathrm d}\ln L_t}.
\label{eq:betafunc}
\end{equation}

In the critical regime, where $L_t\ll\xi$, the SPS hypothesis implies that
\begin{equation}
\ln\Lambda=
F\left( L_t^{1/\nu}\psi\right).
\label{eq:defF}
\end{equation}
Here $\nu$ is the critical exponent that describes the
divergence of the localization length at the critical point,
$F$ is a scaling function, and
\begin{equation}
	\psi\equiv\psi\left(E,W\right)
\label{eq:psi}	
\end{equation}
is a smooth function of disorder and energy that goes to zero linearly
at the critical point.
Equations (\ref{eq:defF}) and (\ref{eq:psi}) are used to fit the results of
numerical simulations for systems in the critical regime.
Once the form of $F$  and the critical exponent $\nu$ are determined, the 
$\beta$-function is calculated by differentiating Eq. (\ref{eq:defF}).

When data outside the critical regime are also included in the analysis, it is 
more practical to use a different form of the SPS law that expresses
$\Lambda$ as a function of the ratio of the
system size to a single relevant length scale $\xi$,
\begin{equation}
	\ln \Lambda=F_{\pm} \left(\frac{L_t}{\xi}\right).
	\label{eq:defFpm}
\end{equation}
The subscript distinguishes the scaling function in the metallic and localized phases.
We follow the convention that $+$ indicates the metallic phase and
$-$ the localized phase.
Data for the metallic and localized phases are analyzed separately.
The $\beta$-function can again be obtained by differentiating Eq. (\ref{eq:defFpm}) once
$F_{\pm}$ have been determined.

We accumulated numerical data for the localized, critical and metallic regimes,
fitted them with the appropriate form.
The best fit to the data was determined by minimizing the $\chi^2$
statistic and
the quality of the fit assessed by the goodness of fit
probability $Q$.
The precision of the results of the fitting procedure were determined using a 
Monte Carlo method\cite{numrep} and expressed as $95\%$ confidence intervals.
We did not include any corrections to SPS due to irrelevant variables\cite{slevin:99} 
since these are negligible for the SU(2) model.\cite{asada:02-1}

\section{Scaling analysis of the renormalized localization length}

\subsection{The critical region}
\label{subsec:critical}

\begin{figure}[tb]
\includegraphics[width=\linewidth]{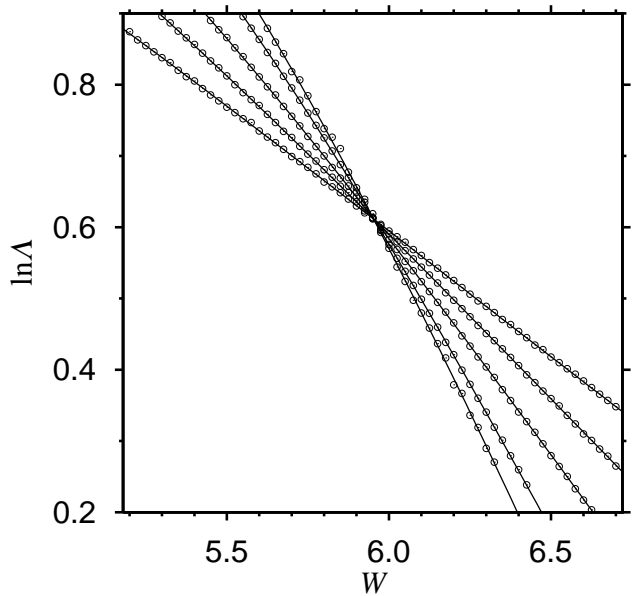}
\caption{A plot of the best fit to the data for the critical regime.
All the curves cross at a common point (the critical point).
This illustrates that the magnitude of any corrections to SPS are smaller 
than the precision of the data. }
\label{fig:critical}
\end{figure}

For the critical region we simulated data with a fixed energy.
When fitting we approximated the function $\psi$
by a  Taylor series truncated at order $n_{\psi}$
\begin{equation}
	\psi=\psi_1 w + \psi_2 w^2 + \cdots + \psi_{n_{\psi}} w^{n_{\psi}},
	\label{eq:psitaylor}
\end{equation}
where
\begin{equation}
	w=\frac{W_c-W}{W_c}
	\label{eq:w}
\end{equation}
Here $W_c\equiv W_c\left(E\right)$ is the critical disorder for the given energy.
The function $F$, which for finite $L_t$ is a smooth function of
energy and disorder, was approximated by a Taylor series truncated at order $n_{0}$
\begin{equation}
	F\left(x\right) = \ln \Lambda_c + x + a_2 x^2 + \cdots + a_{n_{0}} x^{n_0}
	\label{eq:ftaylor}
\end{equation}
The coefficients in both Taylor series, the critical disorder and the critical exponent
are fitting parameters.
The results of the scaling analysis are
given in Table~\ref{table:critical} and the best fit is displayed in Fig.~\ref{fig:critical}.

\begin{table}[tb]
\caption{\label{table:critical}
The results of the scaling analysis in the critical region.
The best fit parameters for energy $E=1$, $L_t=8,16,32,64,96$ and $W\in\left[5.2,6.7\right]$
are listed.
The precision of the data $\Lambda$ is $0.1\%$,
except for $L_t=96$ where it is  $0.4\%$.
Here, and in later tables, $N_d$ is the number of data points,
$N_p$ the number of fitting parameters,
and $Q$ the goodness of fit probability.}
\begin{ruledtabular}
\begin{tabular}{ll|ll}
 $\nu$	& $2.746\pm.009$ & $n_0$  &   3   \\
$W_c$	& $5.953\pm.001$ & $n_{\psi}$&    2   \\
$\ln\Lambda_c$ & $0.6116 \pm .0007$& $N_p$ &  7 \\
$a_2$	& $-0.30 \pm .02$& $N_d$ & $231$   \\
$a_3$	& $-0.01\pm .03$ & $\chi^2$   & $221$   \\
$\psi_1$& $0.986 \pm .004 $ & $Q$ & $0.5$ \\
$\psi_2$& $0.54 \pm .05$ &   &
\end{tabular}
\end{ruledtabular}
\end{table}

\subsection{The insulating phase}
\label{subsec:insulator}

For very strong disorder the localization length $\xi$ is short and
our data satisfy $L_t\gg\xi$. In this limit we expect that
\begin{equation}
	\Lambda \approx \frac{\xi}{L_t}
\end{equation}
This corresponds to the following limiting value of the $\beta$-function
\begin{equation}
	\lim_{\Lambda\rightarrow 0} \beta\left(\ln \Lambda \right)= -1
\end{equation}
Supposing that deviations from this limiting value for small but finite $\Lambda$ are 
of the form
\begin{equation}
	\beta\left(\ln \Lambda \right) = -1 + a \Lambda,
\end{equation}
we arrive at the following form for $\Lambda$
\begin{equation}
\Lambda^{-1}= a+ \frac{L_t}{\xi} \; \; (\Lambda\ll 1)
\label{eq:fitLocAsymp}.
\end{equation}
We found that this fits data for the strongly localized limit well.
The fitting parameters are $\xi\equiv\xi \left(W\right)$ at each disorder $W$,
and $a$.
The details of the best fit obtained with (\ref{eq:fitLocAsymp}) are tabulated in Table \ref{table:insulatorasymptotic}. 

\begin{table}[tb]
\caption{\label{table:insulatorasymptotic}
The scaling analysis for the strongly localized limit.
The ensemble transfer matrix method with $N_r=1000$ and $L_z=1000$ was used.
The best fit to data satisfying the criterion $\Lambda<1/6$ obtained
in simulations of systems with $E=1$ and $L_t\in\left[24,128\right]$ is shown.
The precision of $\Lambda$ is $0.3\%$}
\begin{ruledtabular}
\begin{tabular}{cc|cc}
$W$ & $\xi$ &  &  \\
\hline
10.0 & $10.77 \pm .06$	& $a$  & $1.39\pm.03$ \\
11.0 & $7.12 \pm .03$ &  & \\
12.0 & $5.21 \pm .02$ & $N_d$ 	& 24 \\
12.5 & $4.57 \pm .02$ &	$N_p$  	& 6 \\
13.0 & $4.07 \pm .01$ & $\chi^2$ & 10 \\
     &                & $Q$     & 0.9 \\
\end{tabular}
\end{ruledtabular}
\end{table}

\begin{figure}[tb]
\includegraphics[width=\linewidth]{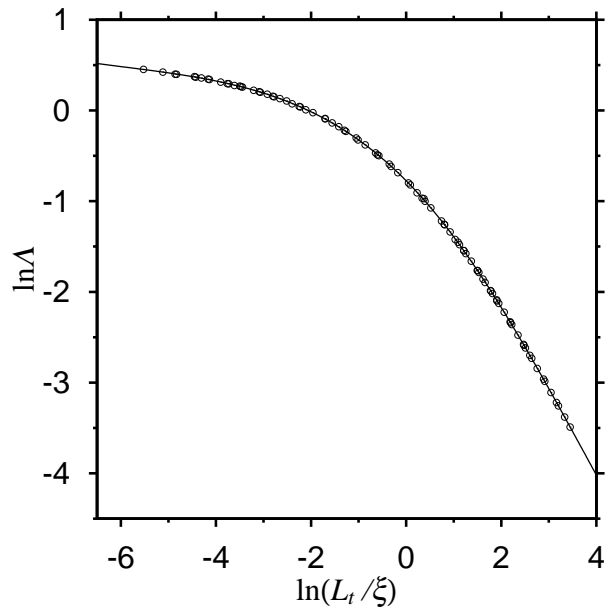}
\caption{
A plot of the best fit to data in the localized phase.
}
\label{fig:insulator}
\end{figure}

To fit all the data in the localized regime we used a cubic spline
parameterization of the scaling function.
We set $x=\ln \left(L_t/\xi\right)$ and $y=\ln \Lambda$ and 
fit the data with $y=f_-\left(x\right)$ where $f_-$ is a cubic spline.
The values of the function $f_-$ at a given set of $x$ values,
the derivatives of $f_-$ at the end points of the
interval considered, and the localization length for each disorder, are fitting parameters.
The subroutines ^^ ^^ spline" and ^^ ^^ splint"
in Ref.~\onlinecite{numrep} are used to perform the
spline interpolation.
To fix the absolute scale of the localization length in this spline fit, we 
set the localization length at $W=12$ to the value obtained 
with (\ref{eq:fitLocAsymp}) in the strongly localized region.
The results are tabulated in Table \ref{table:insulator} and displayed in
Fig \ref{fig:insulator}.

\begin{table}[tb]
\caption{\label{table:insulator}
The fit to data for the localized regime:
the localization length $\xi$ at each disorder $W$
and the parameters for the cubic spline interpolation.
Note that $f_-\left(x\right)\equiv F_-\left(e^{x}\right)$.
The data used are for $E=1$, $L_t\in [16,128]$.
The precision of $\Lambda$ ranges from $0.3\%$ to $1.0\%$.}
\begin{ruledtabular}
\begin{tabular}{cc|cc}
$W$ & $\xi$ &  &  \\
\hline
6.3 & $4006 \pm 200$ & $f_-(-6.5)$ & $0.52 \pm .04$ \\
6.4 & $2061 \pm  80$ & $f_-(-4)$ & $0.325 \pm .004$ \\
6.5 & $1187 \pm  30$ & $f_-(-2)$ & $-0.015 \pm .004$ \\
6.7 & $528  \pm  10$ & $f_-(0)$  & $-0.775 \pm .003$ \\
7.0 & $228  \pm   3$ & $f_-(1)$  & $-1.393 \pm .003$ \\
7.5 & $87.9 \pm .8 $ & $f_-(2)$  & $-2.172 \pm .003$ \\
8.0 & $45.7 \pm .3$  & $f_-(4)$  & $-4.02 \pm .02$ \\
9.0 & $19.06  \pm .09$ & $f_-'(-6.5)$  & $-0.07 \pm .06$ \\
10.0 & $10.76 \pm .04$ & $f_-'(4)$     & $-0.96 \pm .04$  \\
11.0 & $7.12  \pm .03$ &    &  \\
12.0 & $5.21$ (fixed) & $N_d$ & $91$ \\
12.5 & $4.57  \pm .02$ & $N_p$ & $21$ \\
13.0 & $4.07  \pm .02$ & $\chi^2$ & $68$ \\
     &                 & $Q$    & $0.6$ \\
\end{tabular}
\end{ruledtabular}
\end{table}

\subsection{The metallic phase}
\label{subsec:metal}

\begin{table}[tb]
\caption{\label{table:metalasymptotic}
The best fit to data for the strongly metallic limit.
The 
ensemble transfer matrix method with $N_r=10000$ and $L_x=10000$ was used.
The best fit to data satisfying the criterion $\Lambda>4$ obtained
in simulations of systems with $E=1$ and  $L_t \in [16,128]$ is shown.
The precision of $\Lambda$ ranges from $0.3\%$ to $1.0\%$.}
\begin{ruledtabular}
\begin{tabular}{ll|ll}
$W$ & $\xi$ & & \\
\hline
0.0 & $1.69 \pm .07$ & $b$ & $4.48\pm.02$ \\
1.0 & $1.99 \pm .08$ & $c$ & $0.64\pm.01$ \\
2.0 & $3.7  \pm  .1$ & & \\
2.5 & $5.7  \pm  .2$ & $N_d$ & $48$ \\
3.0 & $10$ (fixed)   & $N_p$ & $8$ \\
3.5 & $19.7 \pm .6$  & $\chi^2$ & $44$ \\
4.0 & $45   \pm 2$   & $Q$ & $0.3$ \\
\end{tabular}
\end{ruledtabular}
\end{table}

\begin{figure}[tb]
\includegraphics[width=\linewidth]{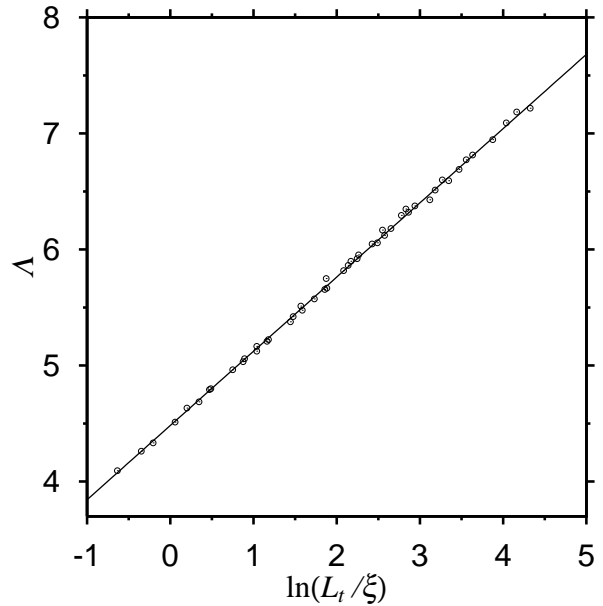}
\caption{
A plot of the best fit to data in the strongly metallic limit.
}
\label{fig:metal2}
\end{figure}

For weak disorder we follow Ref. \onlinecite{mackinnon:83} and argue as follows.
When $\Lambda\gg 1$ there is negligible localization of the electron in the
transverse direction on the quasi-1D system.
As a result the electron sees an effective random potential that is the average
of the random potential in the transverse direction.
In the longitudinal direction the electron is localized with a quasi-1D 
localization length that can be estimated from perturbation theory for a
strictly 1D system.
The result is 
\begin{equation}
	\Lambda \sim \frac{1}{W^2}
\end{equation}
The limiting value of the $\beta$ function that this corresponds to is
\begin{equation}
	\lim_{\Lambda\rightarrow \infty} \beta\left(\ln \Lambda \right)= 0
\end{equation}
For large but finite $\Lambda$ we speculate that deviations from this
can be described by an expansion in powers of $1/\Lambda$. Stopping at the first term 
we have
\begin{equation}
	\beta\left(\ln \Lambda \right) = \frac{c}{\Lambda}.
\end{equation}
This corresponds to  a logarithmic increase of $\Lambda $ with $L_t$
\begin{equation}
\Lambda=b+c\ln \frac{L_t}{\xi} \; \; (\Lambda\gg 1)
\label{eq:fitMetAsymp}
\end{equation}
Data for large $\Lambda$ are well fitted by this form.
Here $b$, $c$ and the correlation length $\xi$ at each disorder $W$,
are fitting parameters.
Since the absolute scale of $\xi$ in the metallic phase
is arbitrary, we set the correlation length at $W=3$ to $\xi(W=3)=10$ to fix the scale.
This does not affect the form of the scaling function or the $\beta$-function.
The best fit is tabulated in Table~\ref{table:metalasymptotic} and displayed in
Fig. \ref{fig:metal2}.

\begin{figure}[tb]
\includegraphics[width=\linewidth]{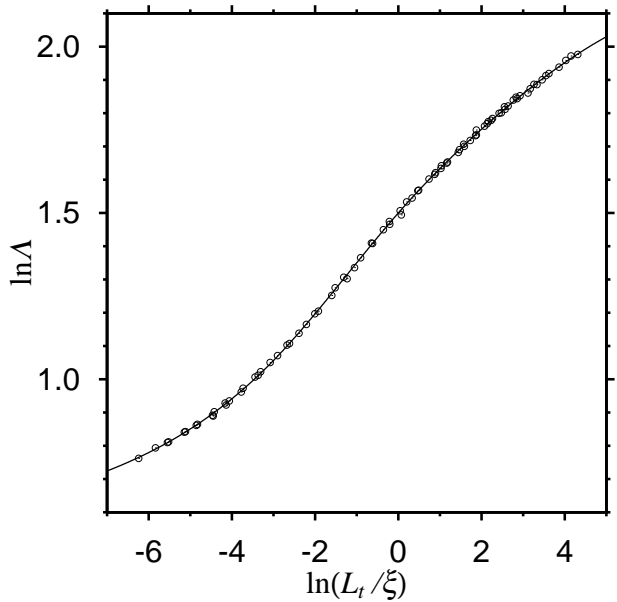}
\caption{
A plot of the best fit to data in the metallic phase.
}
\label{fig:metal}
\end{figure}

To fit data for the whole metallic phase we used a cubic spline interpolation
of the scaling function.
We set $x=\ln\left(L_t /\xi\right)$ and $y=\ln\Lambda$
and fitted the data with $y=f_+\left(x\right)$ where $f_+$ is the cubic spline.
The correlation length $\xi$ at each disorder $W$
and the parameters for the cubic spline interpolation are the fitting parameters.
The best fit is tabulated in Table \ref{table:metal}
and displayed in Fig. \ref{fig:metal}.

\begin{table}[tb]
\caption{\label{table:metal}
The best fit for the metallic phase:
the correlation length $\xi$ at each disorder $W$
and the parameters for the cubic spline interpolation.
Note that $f_+\left(x\right)\equiv F_+\left(e^{x}\right)$. The data used are for
$E=1$, $L_t\in\left[16,128\right]$.
The precision of the data $\Lambda$ ranges from $0.3\%$ to $1.0\%$.}
\begin{ruledtabular}
\begin{tabular}{cc|cc}
$W$ & $\xi$ & &  \\
\hline
0.0 & $1.72 \pm .08$ & $f_+(-7)$ & $0.72 \pm .02$ \\
1.0 & $2.02 \pm .09$ & $f_+(-3)$ & $1.057 \pm .008$ \\
2.0 & $3.7  \pm  .1$ & $f_+(0)$ & $1.500 \pm .005$ \\
2.5 & $5.7  \pm  .2$ & $f_+(2)$ & $1.752 \pm .003$ \\
3.0 & $10$ (fixed)   & $f_+(5)$ & $2.03 \pm .03$ \\
3.5 & $19.6 \pm .6 $ & $f_+'(-7)$ & $0.05 \pm .02$  \\
4.0 & $46   \pm  2 $ & $f_+'(5)$ & $0.07 \pm .04$ \\
4.5 & $119  \pm  5 $ &  &  \\
5.0 & $438  \pm  30$ & $N_d$ & 84 \\
5.3 & $1393 \pm 100$ & $N_p$ & 18 \\
5.5 & $4005 \pm 400$ & $\chi^2$ & 81 \\
5.6 & $8223 \pm 900$ & $Q$ & 0.1  \\
\end{tabular}
\end{ruledtabular}
\end{table}

\section{Estimation of the $\beta$-function}
\label{subsec:betafunc}

\begin{figure}[tb]
\includegraphics[width=\linewidth]{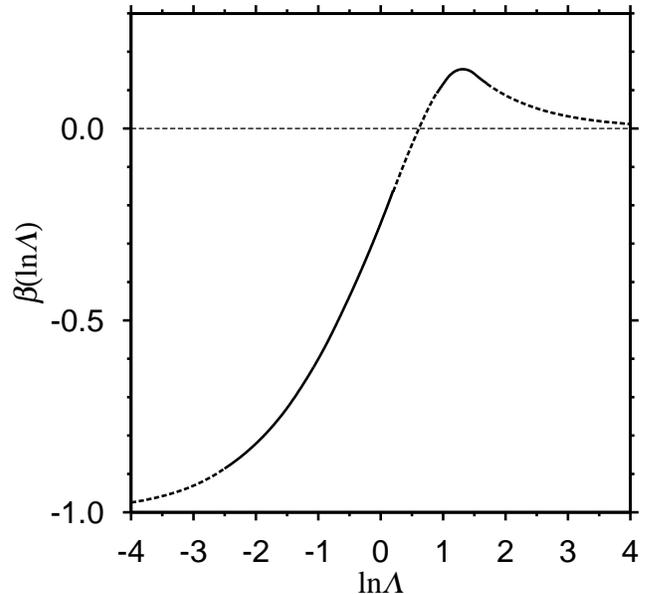}
\caption{
The $\beta$-function for the renormalized localization length.
The different regions (strongly localized, localized, critical, metallic and strongly metallic) are
indicated by the alternating use of solid and dashed lines.
}
\label{fig:betafunc}
\end{figure}

For the critical region, we find after differentiating Eq. (\ref{eq:defF}) 
the following expression for
the $\beta$-function
\begin{equation}
\beta(F\left(s\right))=\frac{1}{\nu} s F^{\prime}\left(s\right) \\
\end{equation}
For the metallic or localized phases, differentiating (\ref{eq:defFpm}) we find
\begin{equation}
\beta(F_{\pm}\left(s\right))= s F_{\pm}^{\prime}(s) \\
\end{equation}
In all cases the $\beta$-function is easily displayed using a parametric plot.

For the strongly localized and strongly metallic limits the appropriate
forms of the $\beta$-function in terms of the fitting paramaters $a$, $b$ and $c$ 
have already been given.

The resulting $\beta$-function is displayed in Fig. \ref{fig:betafunc}.
The precise form of the $\beta$-function depends not only on
the universality class but also on the quantity whose scaling is analyzed.
The $\beta$-function for the  renormalized quasi-1D 
localization length discussed here will differ in detail
from the $\beta$-function for the mean conductance,
or the mean resistance, or the typical conductance etc.\cite{slevin:01-1,slevin:03-1}
It will also differ in detail from the $\beta$-function found in 
renormalization group analyses of field theories of Anderson localization.
\cite{hikami:80,bernreuther:86,wegner:89,hikami:92}
The only common features expected to be shared by all $\beta$-functions for
particular universality class are the existence of a zero, which signals
the existence of a transition, and the slope at the zero, which is 
related to the critical exponent.

\section{the phase diagram of the 2D SU(2) model}

\begin{figure}[tb]
\includegraphics[width=\linewidth]{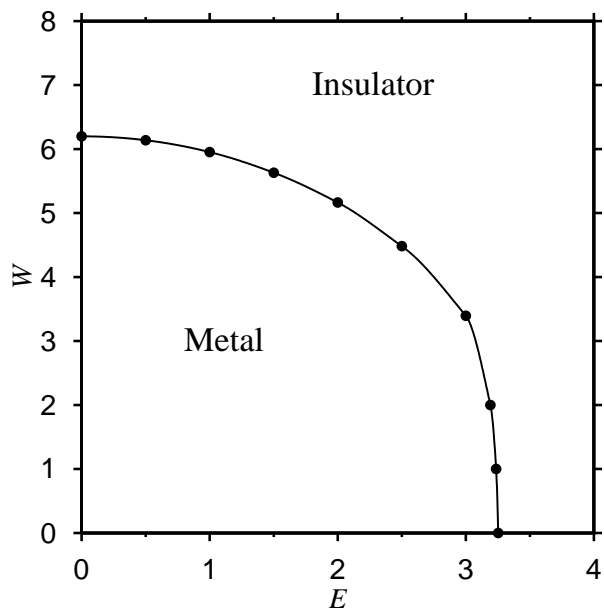}
\caption{
Phase diagram of the SU(2) model.
The line is a cubic spline interpolation.
}
\label{fig:phasediagram}
\end{figure}

The preliminary results for the phase diagram presented in our previous paper left open
the possibility of re-entrant behavior similar to that seen for the Anderson model.
\cite{bulka:87}
To determine whether or not such behavior occurs, the data in our previous
paper were
supplemented by simulations with a fixed disorders
$W=1$ and $W=2$ and varying Fermi energy.
The data were fitted as already described in Sections \ref{sec:fssmethod} 
and \ref{subsec:critical}, the only difference being that in Eq. (\ref{eq:psitaylor}) 
we set
\begin{equation}
	w=\frac{E_c-E}{E_c}
\end{equation}
and determined the critical energy as a function of disorder $E_c\equiv E_c\left(W\right)$.
The results are tabulated in Table \ref{table:phasediagram} and displayed in Fig. \ref{fig:phasediagram}.
Re-entrant behavior is clearly ruled out.

\begin{table}[tb]
\caption{\label{table:phasediagram}
The details of the simulations and fits used to map out the phase
digram of the SU(2) model.}
\begin{ruledtabular}
\begin{tabular}{lllllll}
$E$(fixed) & $L_t$
& $N_{d}$ & $Q$ & $W_c$ & $\ln \Lambda_c$ & $\nu$ \\ \hline
0.0 & [8,64] & 59
& 0.4 & 6.199$\pm$.003 & 0.612$\pm$.002 & 2.75$\pm$.04 \\
0.5 & [8,32] & 51
& 0.5 & 6.139$\pm$.004 & 0.612$\pm$.002 & 2.72$\pm$.04 \\
1.5 & [8,32] & 51
& 0.3 & 5.631$\pm$.004 & 0.611$\pm$.002 & 2.74$\pm$.04 \\
2.0 & [8,64] & 62
& 0.4 & 5.165$\pm$.004 & 0.609$\pm$.002 & 2.73$\pm$.03 \\
2.5 & [16,64] & 47
& 0.1 & 4.483$\pm$.005 & 0.608$\pm$.003 & 2.78$\pm$.05 \\
3.0 &[16,64] & 47
& 0.4 & 3.394$\pm$.006 & 0.611$\pm$.003 & 2.77$\pm$.06 \\
\hline
$W$(fixed) & $L_t$
& $N_d$& $Q$ & $E_c$ & $\ln \Lambda_c$ & $\nu$ \\ \hline
2.0 &[16,64] & 48
& 0.5 & 3.1922$\pm$.0006 & 0.607$\pm$.002 & 2.70$\pm$.04 \\
1.0 &[16,64] & 36
& 0.7 & 3.2367$\pm$.0004 & 0.609$\pm$.003 & 2.70$\pm$.04 \\
0.0 &[16,64] & 31
& 0.7 & 3.2531$\pm$.0003 & 0.613$\pm$.004 & 2.77$\pm$.05 \\
\end{tabular}
\end{ruledtabular}
\end{table}

\section{Summary and Discussion}
In this paper, we
 analyzed the scaling of the renormalized localization length
in the 2D SU(2) model.
We estimated the critical exponent $\nu=2.746\pm 0.009$
and the $\beta$-function.
We also clarified the phase diagram.

The properties of the metallic phase in the 2D symplectic universality class
are of particular interest.
According to the single parameter scaling theory,
this phase has a prefect conductivity.\cite{hikami:80,kawarabayashi:96-2}
This is in spite of the system being disordered.
This conclusion might be avoided if there was some breakdown of single parameter scaling in 
the metallic regime.
However, we have verified clearly in this work that the renormalized localization
length $\Lambda$ does obey the single parameter scaling law 
in the metallic regime.

\appendix

\section{SU(2) gauge transformation}
\label{app:gaugetrans}

Here we describe the SU(2) gauge transformation 
mentioned in the text.
The Lyapunov exponents
are independent of the choice of gauge.

Taking the $x$-direction as the longitudinal direction
and $y$-direction as the transverse direction,
the local SU(2) gauge transformation is given by
\begin{eqnarray}
\left(
\begin{array}{c}
c_{xy\uparrow}\\
c_{xy\downarrow}
\end{array}
\right)
&=&
U(x,y)
\left(
\begin{array}{c}
\tilde{c}_{xy\uparrow}\\
\tilde{c}_{xy\downarrow}\\
\end{array}
\right)
\end{eqnarray}
where $U(x,y)\in$ SU(2) has elements
\begin{eqnarray}
U(x,y)=R(x,y;x-1,y)R(x-1,y;x-2,y) \cdots
\nonumber \\
\cdots R(2,y;1,y) R(1,y;0,y).
\end{eqnarray}
After this transformation, the SU(2) model Hamiltonian has the same form as
Eq. (\ref{eq:hamiltonian}) but with the hopping matrix $R$ replaced with
$\tilde{R}$, where
in the $x$-direction $\tilde{R}(x,y;x+1,y)$ is the unit matrix
and in the $y$-direction
\begin{equation}
\tilde{R}(x,y;x,y+1)=
U(x,y)^{\dagger}
R(x,y;x,y+1)
U(x,y+1).
\end{equation}
The matrix $\tilde{R}(x,y;x,y+1)$ is again uniformly and independently distributed
on SU(2).

\bibliography{su22d}

\end{document}